\documentclass[11pt,%
 amsmath,amssymb,
 aps,
pre,
]{revtex4-2}

\usepackage{graphicx}
\usepackage{dcolumn}
\usepackage{bm}

\begin{document}

\preprint{APS/123-QED}

\title{Aging of glass-forming materials following a temperature jump}

\author{Ayata Ueno}
\affiliation{%
 Graduate School of Science and Technology, Kyoto Institute of Technology, Kyoto 606-8585
}%
\author{Tomoko Mizuguchi}%
 \email{mizuguti@kit.ac.jp}
\affiliation{%
 Faculty of Materials Science and Engineering, Kyoto Institute of Technology, Kyoto 606-8585 
}%
\author{Takashi Odagaki}
\affiliation{
 Research Institute of Science Education, Inc., Kyoto 603-8346
}%

\date{\today}

\begin{abstract}
Physical aging is one of the non-equilibrium phenomena where physical properties change over time due to structural relaxation. Aging in spin glass systems has been explained by a trap model on the temperature-independent energy landscape. Meanwhile, in the free energy landscape (FEL) approach to aging phenomena, it is assumed that the FEL responds to temperature changes with a time delay. 
In this paper, aging in a glass forming model in which both the trapping effect and the delayed response of the FEL exist is studied after the temperature is changed. It is confirmed that the trapping effect gives rise to Type-I aging where the relaxation time increases with waiting time regardless of the direction of temperature change, and that the delayed response of the FEL produces Type-II aging where the waiting-time dependence of the relaxation time depends on the direction of temperature change. When both effects exist and the response time of the FEL is appropriate, these effects can be differentiated in the short-time behavior of the temporal relaxation time. It is argued that the material time or the internal clock and the fictive temperature introduced phenomenologically are understood as the concepts describing the delayed response of the FEL to temperature change.
\end{abstract}

\maketitle


\section{Introduction}
Nonequilibrium systems such as glass-forming materials exhibit properties completely different from those observed in equilibrium systems, where the physical properties observed in experiments after a perturbation is applied depend on the time when the observation is made or started. This phenomenon is called aging~\cite{Bouchaud1997,Vincent1997,Berthier2011,Micoulaut2016}. 
Aging has been observed for many physical properties including enthalpy relaxation~\cite{Cowie1989,Hutch1999,Cangialosi2013,Sakatsuji2016}, dielectric response~\cite{Fukao2009,Hecksher2010,Wojnarowska2021}, intermediate scattering function~\cite{Kob2000-2} in various kinds of systems including spin glasses~\cite{Lundgren1983,Ocio1985,Vincent1986}, structural glasses~\cite{Sciortino2002,Kob2000-1}, and polymers~\cite{White2006,McKenna2017}. 

Most common experiment is an observation of response to a temperature change~\cite{Niss2017,Hecksher2019,Riechers2022}. In this experiment, the temperature of heat bath is raised ($T$-up protocol) or lowered ($T$-down protocol) at time $t=0$ and physical properties are measured. The physical properties age and relax to new equilibrium values. Measurement is sometimes made after a waiting time $t_w$, and the waiting time dependence of the relaxation is investigated.

Aging is related to the slow dynamics. In fact, the slow dynamics is often represented by the Kohlrausch-Williams-Watts (KWW) relaxation function $\phi_{\mathrm{KWW}}(t)=\exp[-(t/\tau)^{\beta}]$ ($0<\beta<1$) whose temporal relaxation time defined by $\tau_{tmp}=[-\partial\ln(\phi_{\mathrm{KWW}}(t))/\partial t]^{-1}$ is an increasing function of the observation time. The temporal relaxation time is identical to the inverse of the Kovacs-McKenna relaxation rate~\cite{Kovacs1963,McKenna1994}. We can define two-time relaxation function $\phi(t^{\prime},t_w)$ associated with relaxation function $\phi(t)$ by $\phi(t^{\prime},t_w)=\phi(t^{\prime}+t_w)/\phi(t_w)$. It is straightforward to show that the relaxation time of the KWW two-time relaxation function $\phi_{\mathrm{KWW}}(t^{\prime},t_w)$ as a function of $t^{\prime}$ is an increasing function of $t_w$. Let us consider a system which shows an aging due to the slow relaxation represented by the KWW function. When the temperature is raised or lowered at time $t=0$, physical quantities relax to new equilibrium values and the relaxation function shows aging in which the relaxation time always increases with $t_w$ regardless of the direction of temperature change. This aging is named as Type-I aging. The origin of the slow relaxation has been explained by the trap model~\cite{Bouchaud1992}, where the escape rate from a trap obeys a power law distribution with a negative power and the system may find a deeper trap as the time passes.

Aging has been studied on the basis of energy landscape models. 
The energy landscape approach for glassy systems was first proposed by Stillinger and Weber in the 1980s as the potential energy landscape~\cite{Stillinger1982,Stillinger1983}. Later, energy landscape pictures have extended to an enthalpy landscape~\cite{Mauro2012,Mauro2021} and a free energy landscape (FEL)~\cite{Odagaki2017}. These approaches succeeded in describing relaxation processes such as a crystal growth rate~\cite{Mauro2021} and a non-linear dielectric response~\cite{Odagaki2012}.
Diezemann focused on the difference between translational and rotational relaxation and assumed that the FEL responds to the temperature change without delay to explore the aging properties~\cite{Diezemann2005}.

In experiments on the dielectric loss reported by Hecksher et al.~\cite{Hecksher2010}, the temporal relaxation time becomes longer with time regardless of the direction of temperature change. Although this aging may be of Type-I, we need more careful analysis to judge it as we will discuss later. Riechers et al.~\cite{Riechers2022} showed that the relaxation of the potential energy after a temperature change becomes faster with time for $T$-up protocol and slower with time for $T$-down protocol in molecular dynamics (MD) simulations for a model glass former. In temperature cycling experiments on a spin glass, Lederman et al.~\cite{Lederman1991} showed that the relaxation time of the thermoremanent magnetization increases with waiting time for both protocols.

Aging has been phenomenologically understood by considering a material time~\cite{Hecksher2019,Riechers2022} or an internal clock~\cite{Hecksher2010} which explain slow dynamics, where these times pass slower than the actual time. The physical meaning of these concepts are still to be clarified.

In explaining the enthalpy relaxation which shows aging, Tool-Narayanaswamy-Moynihan (TNM) introduced a fictive temperature~\cite{Tool1946, Narayana1971, Moynihan1976}, where the instantaneous structure of a glass in non-equilibrium state below $T_g$ is assumed to be the structure of a super-cooled liquid in equilibrium at a fictive temperature. It has been widely used in the analysis of experimental data, allowing quantitative agreement with experimental results on the enthalpy relaxation~\cite{Ribelles1997,Boucher2011,Malek2023}.

Aging following temperature change has been studied by MD simulations for polymers: for some materials, it has been shown that the relaxation time increases with waiting time when the temperature is reduced~\cite{Kob2000-1}. However, computer simulation for a model polymer exhibits aging in which the relaxation time decreases with the waiting time when the temperature is raised to moderately high temperature~\cite{Suarez2009}.

A recent theoretical study on a random walk with delayed response of the jump rate to the temperature change shows that the relaxation time can be an increasing or decreasing function of the waiting time depending on the direction of temperature change~\cite{Odagaki2023}. This aging is classified as Type-II. The delayed response of the jump rate is considered to be a manifestation of the delayed response of the FEL which determines the slow dynamics. 

A key question is then what aging will be observed when both effects, i.e. the trapping mechanism and the delayed response of the FEL, exist. It is also interesting to investigate if various concepts introduced to understand aging can be explained in a unified manner by a solid foundation. To this end, we investigate a trapping diffusion model (TDM) incorporated with the delayed response of the FEL. In Section 2, we explain the model system and procedure for our analysis. We first analyze two effects separately to test our numerical method: in Section 3.1, we confirm that the simple trapping diffusion model produces Type-I aging, and in Section 3.2, we show Type-II aging appears in the regular trapping random walk with a delayed response of the trapping rate. This model can be solved rigorously as presented in Appendix B. In Section 4, we study the extended trapping diffusion model in which both effects are present. We show that when the relaxation of the FEL is not too slow nor too fast, the clear effect appears in the temporal relaxation time. Section 5 is devoted to discussion.

\section{Methods}
\subsection{Model}
In the FEL picture for non-equilibrium systems, the structural relaxation is described by a jump motion of a representative point among basins in the FEL~\cite{Odagaki2017} and the jump rate has been shown to obey power-law distribution in general~\cite{Odagaki1995}. The random motion is described by the trapping diffusion equation where the jump rate is determined by the starting basin alone. The rate $W_n$ that a representative point jumps from a basin $n$ to an adjacent basin in unit time obeys the power-law distribution,
\begin{equation}
F(W_n)=\Bigg\{
\begin{aligned}
\frac{\rho +1}{w_0}&\left(\frac{W_n}{w_0}\right)^{\rho} &(W_n \leq w_0) \\
&0 &(W_n>w_0)
\end{aligned}
\end{equation}
\begin{equation}
\rho=\frac{T-T_g}{T_g-T_K} 
\end{equation}
where $w_0$ is an attempt frequency, and $T_g$ and $T_K$ are the glass transition temperature and the Kauzmann temperature, respectively.
A change in $\rho$ derives a change in relaxation dynamics. $\rho=\infty$ corresponds to the simple Debye relaxation~\cite{Odagaki1990}.

In order to control the temperature and the depth of basins independently, we introduce a model equivalent to Eqs.~(1) and (2) as follows~\cite{Odagaki2012}: we first set
\begin{equation}
W_n = w_0 \exp \left( -\epsilon_n \frac{T_g-T_K}{T-T_K} \right)
\end{equation}
and parameter $\epsilon_n$ representing the inherent depth of basin $n$ obeys the exponential distribution
\begin{equation}
F(\epsilon_n)=\exp(-\epsilon_n)\quad(0 \leq \epsilon_n \leq \infty).
\end{equation}
Note that Eqs.~(3) and (4) produce the power-law distribution of Eqs.~(1) and (2). The jump rate between basins is changed with temperature because the barrier height of FEL changed with temperature, unlike PEL. A model FEL was calculated by using a density functional theory (DFT)~\cite{Yoshidome2006}. In the framework of DFT, the temperature dependence of FEL is derived from the change in the density distribution. 

In order to incorporate the delayed response of the FEL to change in the heat bath temperature, we introduce an internal temperature $T_{\mathrm{int}}(t)$ which represents the change of depths of the FEL in response to the heat bath temperature. Suppose the temperature of the heat bath is changed from $T_i$ to $T_f$ at $t=0$, then, $T_{\mathrm{int}}(t)$ is assumed to show a delayed response such that
\begin{equation}
T_{\mathrm{int}}(t)=T_f+(T_i-T_f)\varphi(t) ,
\end{equation}
where the delayed-response function $\varphi(t)$ of the internal temperature is a monotonically decreasing function satisfying
\begin{equation}
\varphi(t)=\bigg\{
\begin{aligned}
&1 &(t=0) \\
&0 &(t=\infty)  .
\end{aligned}
\end{equation}
Using $T_{\mathrm{int}}(t)$, $W_n$ is assumed to be given by
\begin{equation}
W_n = w_0 \exp \left( -\epsilon_n \frac{T_g-T_K}{T_{\mathrm{int}}(t)-T_K} \right).
\end{equation}
In this approach, the depth and jump rate of all basins change uniformly keeping their relative magnitude to other basins when the temperature is changed (Fig. 1). In the present study, the delayed-response function $\varphi(t)$ is assumed to be an exponential function
\begin{equation}\label{delayed}
\varphi(t)=\exp(-t/\tau_F)
\end{equation}
where $\tau_F$ is the relaxation time of $\varphi(t)$ and represents the resistance of the FEL shape to deformation in response to change in the heat bath temperature. The relaxation time $\tau_F$ is a parameter characterizing the substance which can be determined by experiments on aging. 

\begin{figure}
\includegraphics[width=10cm]{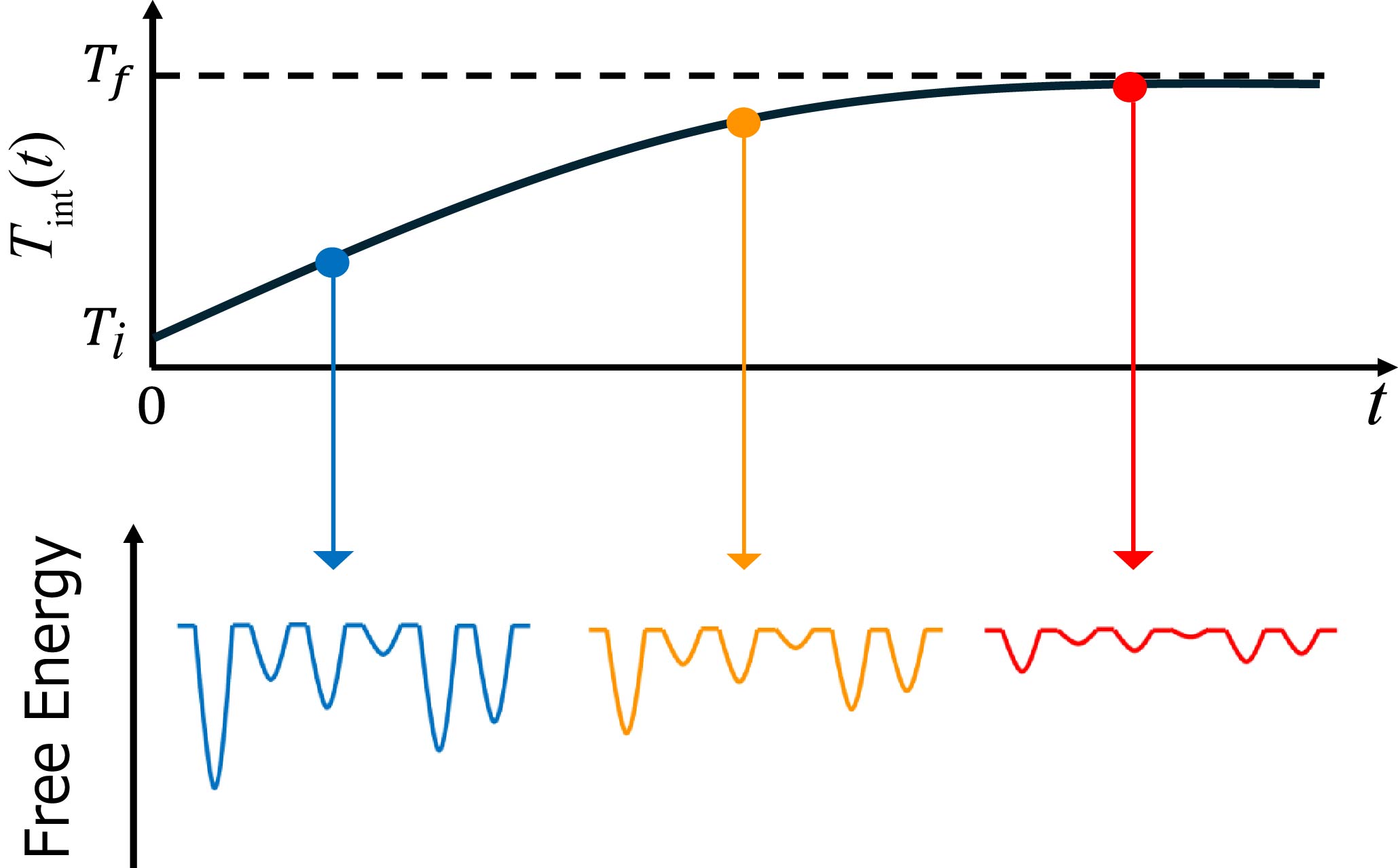}
\caption{Schematic diagram of the time course of the FEL and the internal temperature $T_{\mathrm{int}}(t)$.  As $T_{\mathrm{int}}(t)$ increases, the depth of basins become shallower uniformly leading to weaker trappings and larger jump rates.}
\end{figure}

\subsection{Calculation of the self-intermediate scattering function}
In the FEL approach, the time evolution of a system is represented by a trajectory of the representative point in the FEL, which forms a one dimensional string (reaction coordinate). If we focus on a particular atom and follow its dynamics along one axis in the real space, the dynamics becomes a random walk on the axis. In this paper, we are interested in the relaxation and not in the configurational effects, we consider a trapping random walk on one-dimensional lattice which is described by the master equation~\cite{Odagaki1990}
\begin{equation}
\frac{\partial P(n,t)}{\partial t}=W_{n+1}(t)P(n+1,t)+W_{n-1}(t)P(n-1,t)-2W_{n}(t)P(n,t)
\end{equation}
where $P(n,t)$ is the probability of a representative point being in a basin $n$ at time $t$ and $n$'s constitute a one-dimensional lattice whose lattice constant is $a$.

In this paper, the self-intermediate scattering function (SISF) $F_s (k,t)$ is calculated by the following procedure. We first create a random distribution of ${\epsilon_n}$ and assume the temperature of the heat bath is changed from $T_i$ to $T_f$ at $t=0$. Then, we solve Eq. (9) numerically to obtain $P(n,t)$ under the initial condition that a representative point is in a basin $n_0$ at $t=0$:
\begin{equation}
P(n,t=0)=\delta_{n n_0}.
\end{equation}
$P(n,t)$ obtained under this initial condition is denoted as $P(n,t|n_0,t=0)$. Then, $F_s(k,t)$ is obtained by the spatial Fourier transform of $P(n,t|n_0,t=0)$,
\begin{equation}
F_s(k,t)=\sum_n P(n,t|n_0,t=0) \exp (ik(n-n_0)a)   
\end{equation}
where the wave number $k$ is set to $k^* \equiv ka = 0.79$ in this study. This is $F_s(k,t)$ for an observation started at $t=0$. Note that $P(n,t|n_0,t=0)$ and $F_s(k,t)$ depend on distribution of $\epsilon_n$. Let $F_s(k,t)$ in a given $\epsilon_n$-distribution $s$ be written as $F_s(k,t;s)$. We calculated $F_s(k,t;s)$ for 30 samples with different sets of $\epsilon_n$. Then, $F_s(k,t;s)$ was averaged over $s$ to obtain the result $\langle F_s (k,t;s)\rangle_s$:
\begin{equation}
\langle F_s (k,t;s)\rangle_s=\sum_s P_{\mathrm{std}} (n_0,t=0;s)  F_s (k,t;s)
\end{equation}
\begin{equation}
P_{\mathrm{std}} (n_0,t=0;s) = \frac{W_{n_0} (t=0;s)^{-1}}{\sum_{n_0} W_{n_0} (t=0;s)^{-1}}
\end{equation}
where $P_{\mathrm{std}} (n_0,t=0;s)$ is the probability that a representative point exists in a basin $n_0$ at $t=0$ in a sample $s$.  Here, we assumed that the system is in the steady state at $t=0$ where the probability distribution is that in the steady state at $T_{\mathrm{int}} (t=0)= T_i$~\cite{Haus1982}. The waiting time dependence of $\langle F_s (k,t;s) \rangle_s$ is represented by a two-time relaxation function (TTRF) defined by
\begin{equation}
F_s (k,t^{\prime},t_w )=\frac{\langle F_s (k,t^{\prime}+t_w;s) \rangle_s}{\langle F_s (k,t_w;s) \rangle_s} 
\end{equation}
where $t^{\prime}=t-t_w$ is the elapsed time from $t=t_w$. We discuss aging by $t_w$ dependence of the relaxation time of this function. We also define the temporal relaxation time for TTRF $\tau_{\mathrm{tmp}} (t^{\prime},t_w)$ by
\begin{equation}
\tau_{\mathrm{tmp}} (t^{\prime},t_w)=\Big[ -\frac{\partial \log F_s(k,t^{\prime},t_w)}{\partial t^{\prime}} \Big]^{-1}.
\end{equation}

In the present study, we use scaled parameters denoted with an asterisk, setting the time scale as $w_0^{-1}$ and the distance scale as $a$. We also use $T_K$ as the unit of temperature. In our numerical calculation, we used $\Delta t^*=w_0 \Delta t=0.001$ steps$^{-1}$ for discretization of Eq. (9). In our study, the glass transition temperature is set to $T_g^*=T_g/T_K=1.25$ and the following two temperature sets of $T_i^*$ and $T_f^*$ are used: ($T_i^*,T_f^*$) = ($T_i/T_K,T_f/T_K$) = (1.15,1.24) for $T$-up protocol and (1.24,1.15) for $T$-down protocol. For $\tau_F$, we studied five values: $\tau_F^*=w_0 \tau_F=0,2,5,10$ and 30, where $\tau_F^*=0$ corresponds to the FEL without delay, that is $T_{\mathrm{int}}^*$ changes instantly from $T^*_i$ to $T^*_f$ at $t^*=0$.

All variables and parameters are summarized in Table~\ref{tab1}. Data and program codes that support the findings of this article are openly available~\cite{dataset}.

\begin{table}[]
    \centering
    \begin{tabular}{|c|c|}\hline
         $W_n$ & a jump rate of basin $n$ defined by Eq.~(7)\\ \hline
         $w_0$ & an attempt frequency \\ \hline
         $T_g$ & the glass transition temperature \\ \hline
         $T_K$ & the Kauzmann temperature \\ \hline
         $\epsilon_n$ & the inherent depth of basin $n$ \\ \hline
         $T_{\rm{int}}$ & the internal temperature defined by Eq.~(5)\\ \hline
         $T_i$ & the initial temperature of temperature jump \\ \hline
         $T_f$ & the final temperature of temperature jump \\ \hline
         $\tau_F$ & a relaxation time of the internal temperature \\ \hline
         $a$ & a lattice constant \\ \hline
         $n_0$ & a basin which a representative point exists at $t=0$ \\ \hline
         $s$ & a sample with a given $\epsilon_n$-distribution \\ \hline
         $k$ & the wave number \\ \hline
         $t_w$ & a waiting time before measurement starts \\ \hline
         $t^{\prime}$ & an elapsed time after measurement starts\\ \hline
         $\tau_{\rm{tmp}}$ & the temporal relaxation time for the two-time relaxation function defined by Eq.~(15)\\ \hline
    \end{tabular}
    \caption{All variables and parameters}
    \label{tab1}
\end{table}

\section{Type-I and Type-II agings}
We first investigate two aging effects, the trapping mechanism and the delayed response of the FEL, separately. By setting $\tau_F^*=0$, we investigate the aging in the trapping diffusion model without delayed response of the FEL (this case is called simply the trapping diffusion model hereafter) and obtain the waiting time dependence of TTRF which is shown in Fig. 2 for (a) $T$-down and (b) $T$-up protocol. Apparently, the relaxation time for both protocols increase with the waiting time and the aging is turned out to be of Type-I. It should be remarked that, as shown in Appendix A, the relaxation of SISF can be fitted well by the KWW function only in a limited range of time. In Fig. 3, we show the temporal relaxation time $\tau_{\mathrm{tmp}}^* (t^{\prime *},t^*_w )$ at $t_w^*=0$. The temporal relaxation time is a monotone increasing function for both $T$-down and $T$-up protocols. Figures 2 and 3 show clearly that the trapping diffusion model gives rise to Type-I aging.

\begin{figure}
\includegraphics[width=16cm]{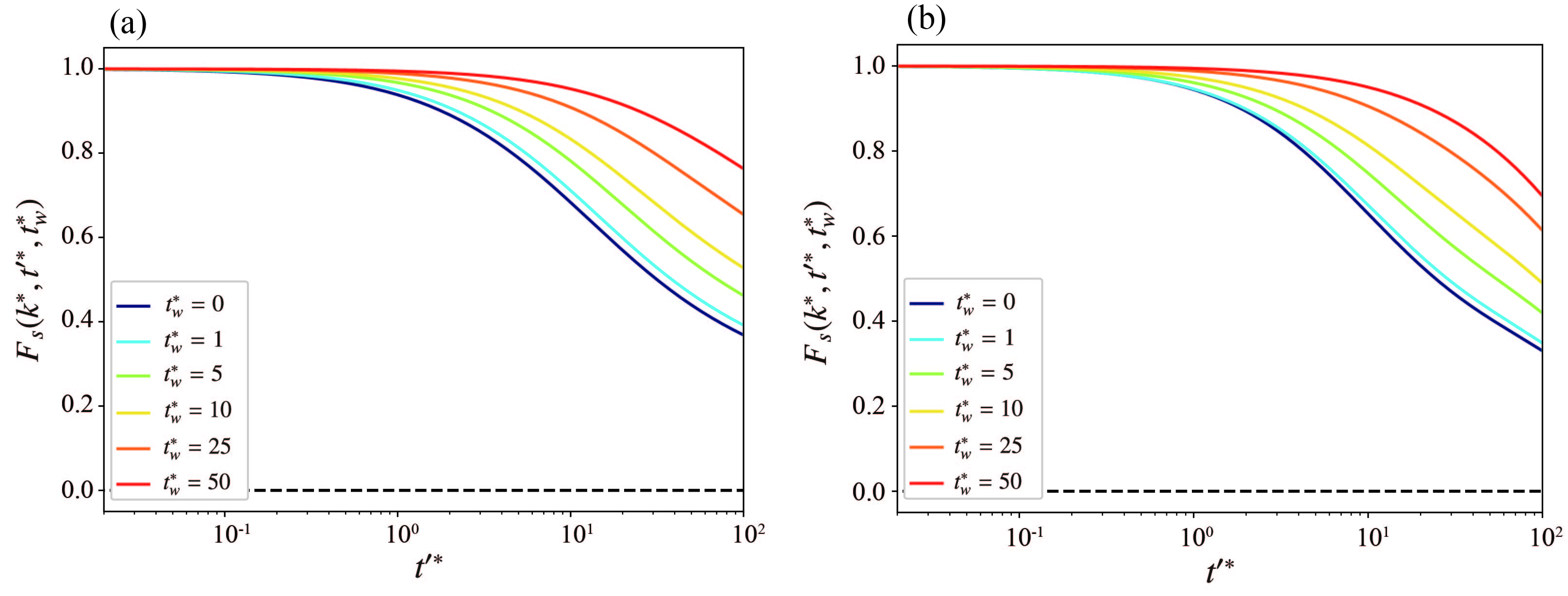}
\caption{Waiting time dependence of TTRF for the trapping diffusion model: (a) $T$-down protocol and (b) $T$-up protocol.}
\end{figure}

\begin{figure}
\includegraphics[width=16cm]{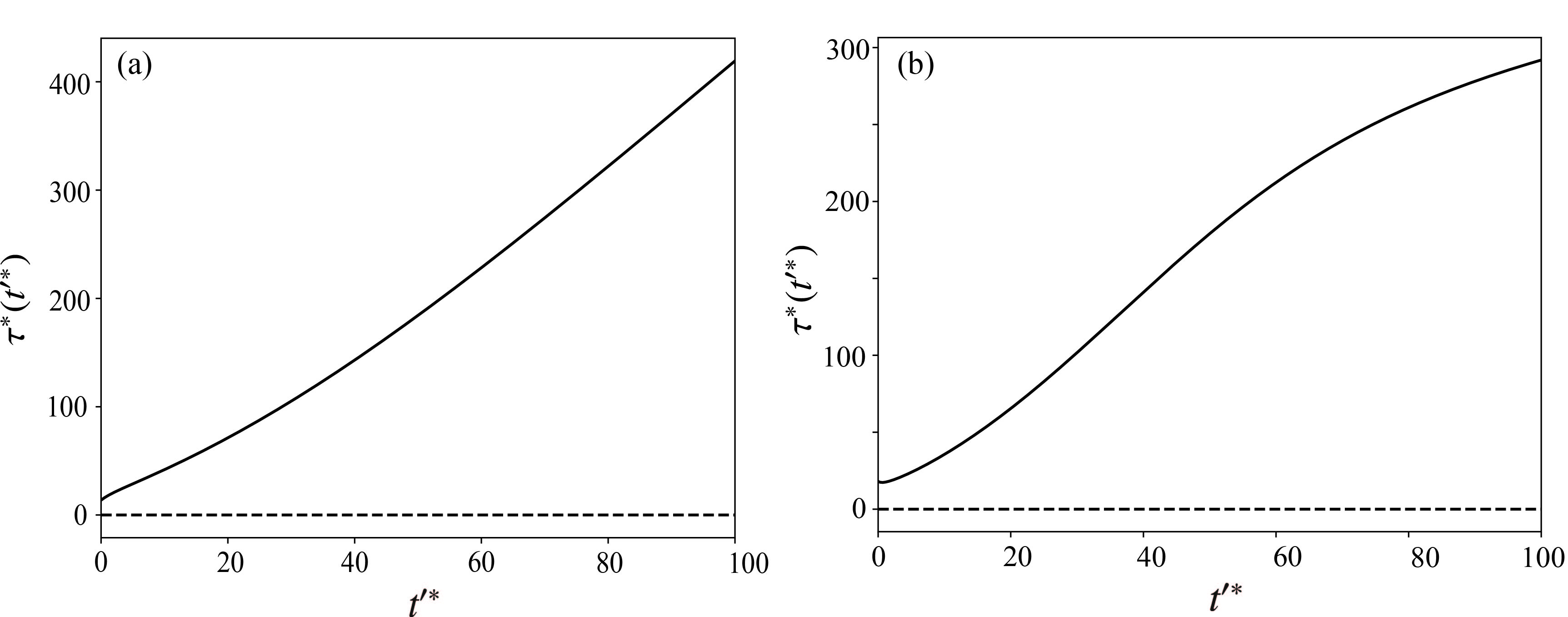}
\caption{Temporal relaxation time $\tau_{\mathrm{tmp}}^* (t^{\prime *},t_w )$ at $t_w^*=0$ of TTRF for the trapping diffusion model: (a) $T$-down protocol and (b) $T$-up protocol.}
\end{figure}

Next, we investigate effects of the delayed response of the FEL alone. To this end, we set $\epsilon_n=1$ for all $n$ (this case is called the trapping random walk hereafter for simplicity). Figure 4 shows the waiting time dependence of TTRF: (a) $T$-down and (b) $T$-up protocol. Apparently, the relaxation time becomes longer in (a) and shorter in (b) with waiting time and therefore the aging is of Type-II. The temporal relaxation time for $T$-up protocol confirms this observation as shown in Fig. 5. The relaxation time becomes shorter with $t^{\prime *}$.

\begin{figure}
\includegraphics[width=16cm]{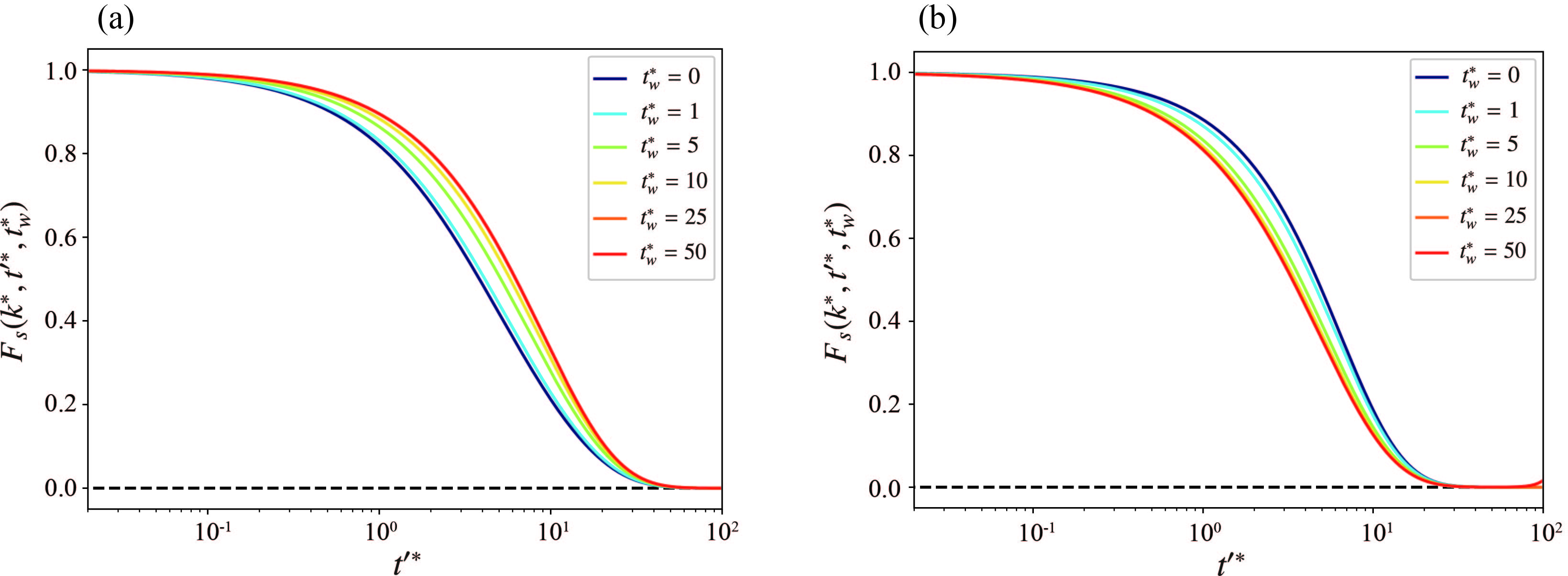}
\caption{Waiting time dependence of TTRF for the trapping random walk with the delayed response of the FEL: (a) For $T$-down protocol, relaxation time becomes longer and (b) for $T$-up protocol, relaxation time becomes shorter with waiting time.}
\end{figure}

\begin{figure}
\includegraphics[width=9cm]{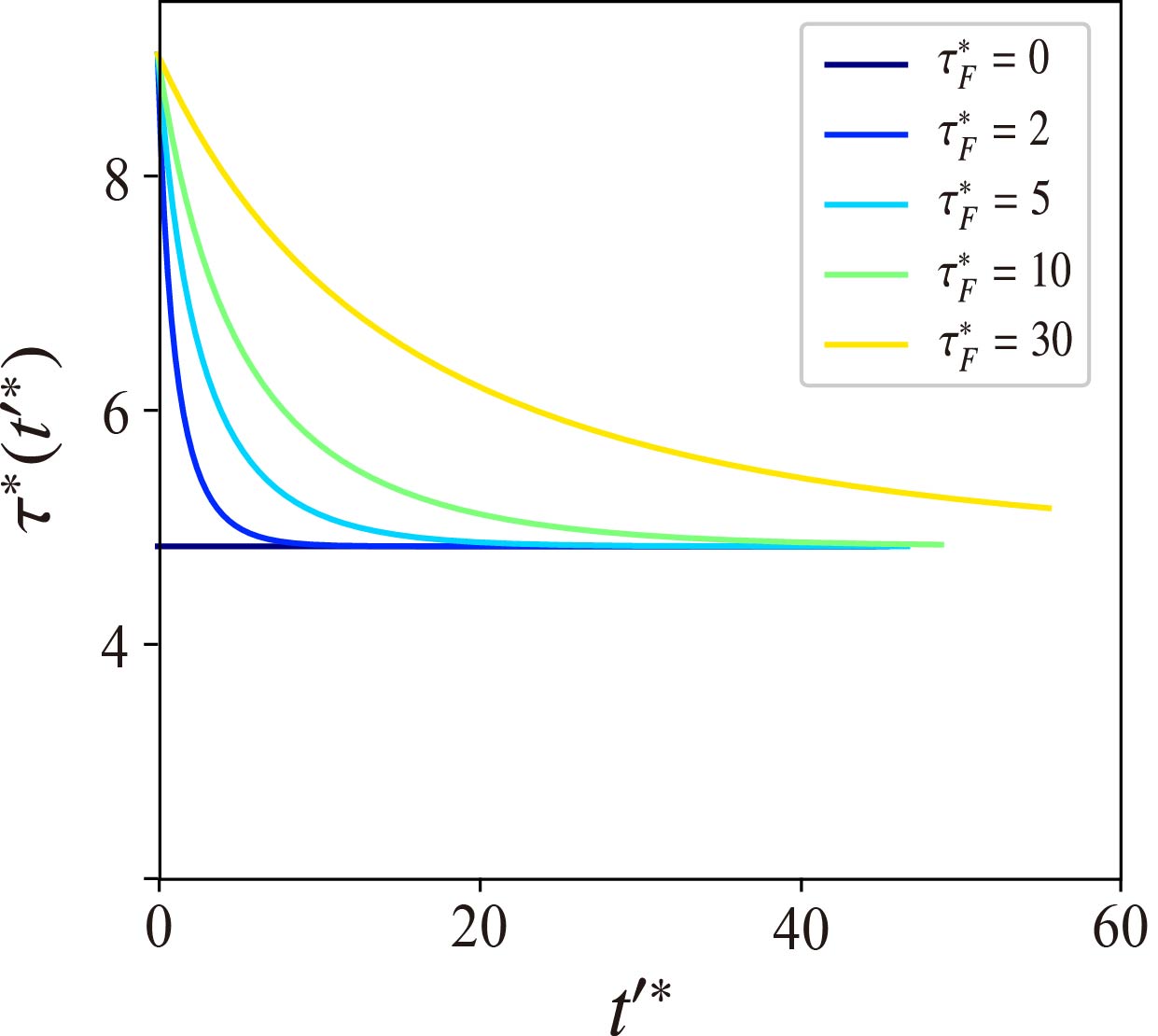}
\caption{The $t^{\prime *}$ dependence of the temporal relaxation time at $t^*_w=0$ of TTRF for the trapping random walk with the delayed response of the FEL for $T$-up protocol.}
\end{figure}

\section{Aging of the extended trapping diffusion model}
Now, we investigate the aging of the model when both effects exist (this case is called an extended trapping diffusion model hereafter). Figure 6 shows $t^{\prime *}$ dependence of TTRF for several $t_w^*$ when $\tau_F^*=5$ : (a) $T$-down protocol and (b) $T$-up protocol. It is seen that the relaxation time of the TTRF increases with waiting time. It is confirmed that this trend does not depend on $\tau_F^*$. As we discuss below, this trend does not indicate the aging is a pure Type-I.

\begin{figure}
\includegraphics[width=16cm]{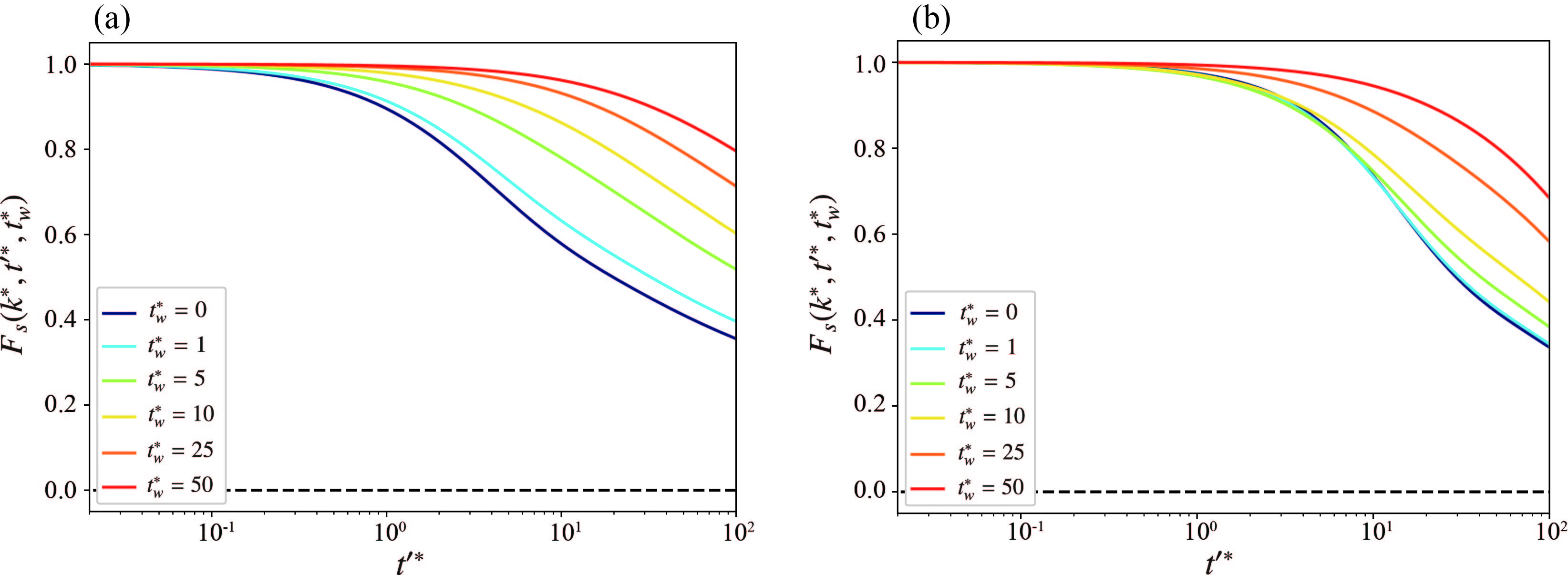}
\caption{$t^{\prime *}$ dependence of TTRF when $\tau_F^*=5$ for the extended TDM: (a) $T$-down protocol and (b) $T$-up protocol.}
\end{figure}

Figure 7 shows the $t^{\prime *}$ dependence of the temporal relaxation time $\tau_{\mathrm{tmp}}^* (t^{\prime *},t^*_w )$ at $t_w^*=0$ when $\tau_F^*=5$: (a) $T$-down protocol and (b) $T$-up protocol. We see that while the temporal relaxation time is a monotone increasing function of $t^{\prime *}$ for $T$-down protocol, it becomes non-monotone function of $t^{\prime *}$ for $T$-up protocol. The initial decay of $\tau_{\mathrm{tmp}}^* (t^{\prime *},0)$ is due to the delayed response of the FEL as shown in Fig. 5 which competes with the increase of the relaxation time due to the trapping mechanism (Fig. 3 (b)) to produce a minimum. In fact, the position of the minimum $t^{\prime *}\sim 6$  can be estimated as the crossing point between the temporal relaxation time of the trapping random walk and that of the trapping diffusion model approximated by $\langle W_n^{-1} \rangle^{-1}$. This indicates that if the initial decay of the relaxation time becomes smaller as $\tau_F^*$ is increased, the minimum in the $\tau_{\mathrm{tmp}}^* (t^{\prime *},0)$ vs $t^{\prime *}$ plot becomes shallower and eventually disappears. The minimum should also disappear when $\tau_F^*=0$ since the temporal relaxation time is a constant. In fact, Fig. 8 shows the $\tau_{\mathrm{tmp}}^* (t^{\prime *},0)$ vs $t^{\prime *}$ plot for various values of $\tau_F^*$ and the minimum disappears for $\tau_F^*=30$ and $\tau_F^*=0$. It is interesting to note that two mechanisms of aging can be identified by $t^{\prime *}$ dependence of the temporal relaxation time for $T$-up protocol.

\begin{figure}
\includegraphics[width=16cm]{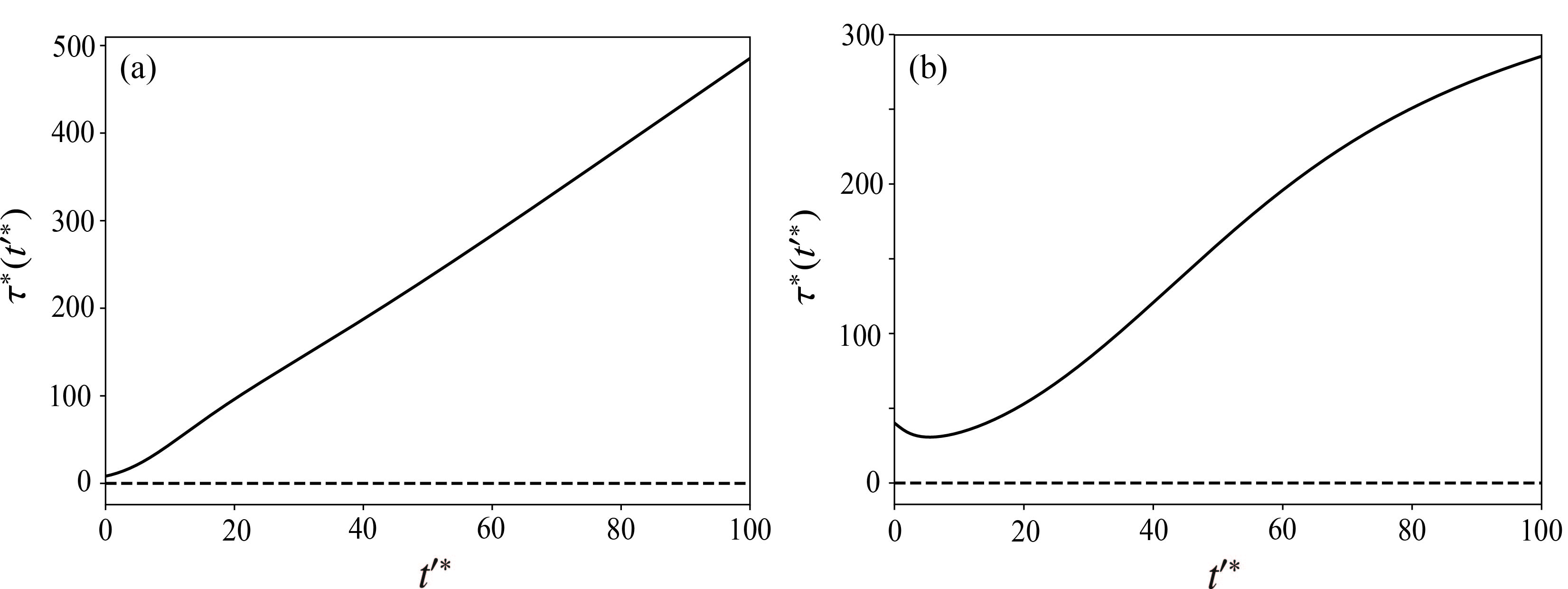}
\caption{$t'^*$ dependence of the temporal relaxation time $\tau_{\mathrm{tmp}}^* (t^{\prime *},t_w^* )$ for $t_w^*=0$ when $\tau_F^*=5$ for the extended TDM model: (a) $T$-down protocol and (b) $T$-up protocol.}
\end{figure}

\begin{figure}
\includegraphics[width=10cm]{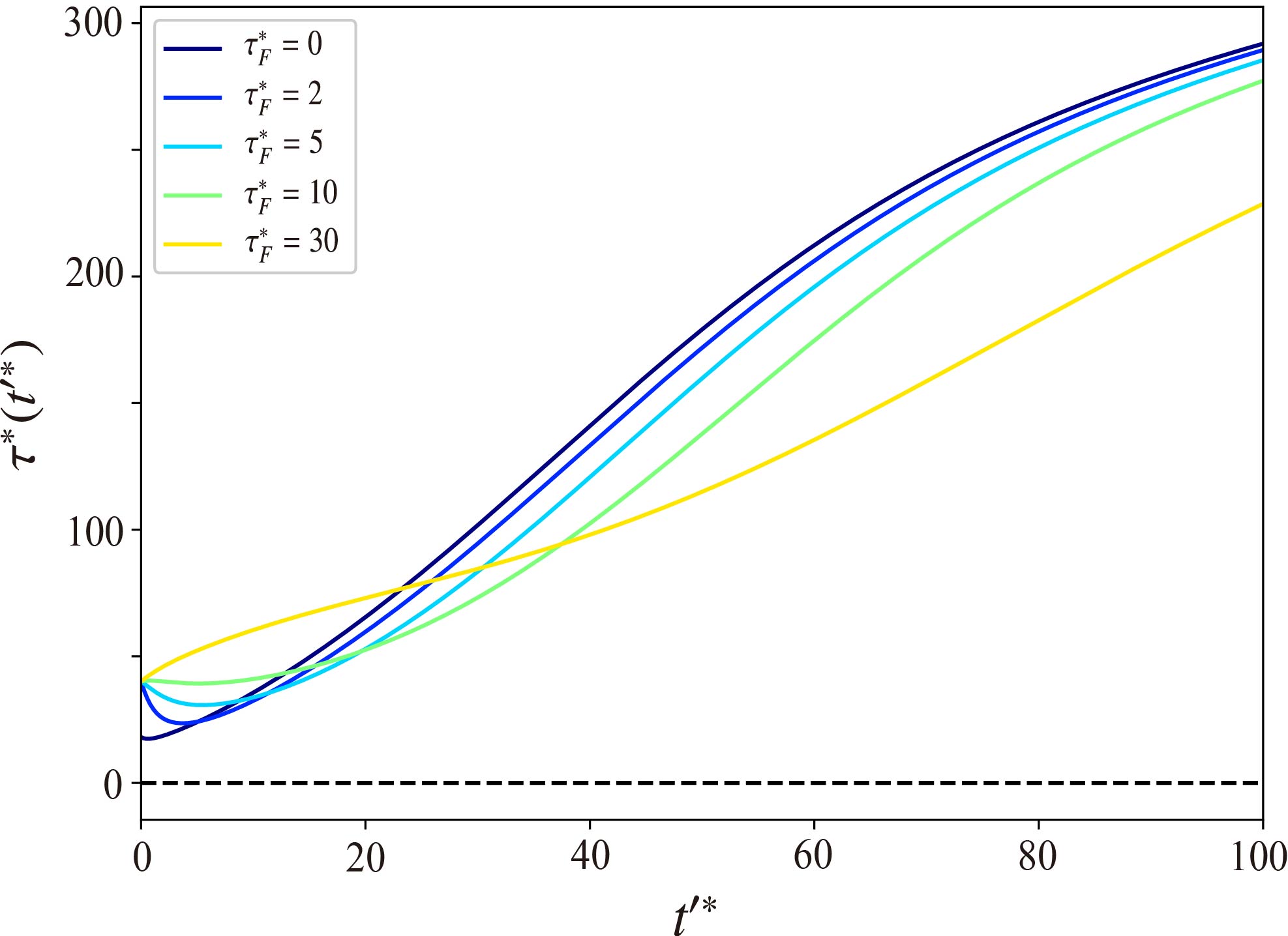}
\caption{$t'^*$ dependence of the temporal relaxation time $\tau_{\mathrm{tmp}}^* (t^{\prime *}, t_w^*)$ at $t_w^*=0$ for $\tau_F^*= 0,2,5,10$ and 30 in $T$-up protocol for the extended TDM.}
\end{figure}

\section{Discussion}
Exploiting numerical simulation, we studied physical aging of a model glass former which has two competing aging mechanisms, the trapping effect and the delayed response of the FEL. These mechanisms are known to produce physical aging in which relaxation time for $T$-up protocol becomes longer in the former and shorter in the latter with waiting time. We showed that for $T$-up protocol of temperature change, the temporal relaxation time may have a minimum as a function of waiting time when the relaxation of the FEL is appropriate. This will explain observation in experiments and simulations fot $T$-up protocol which show that the the relaxation becomes faster with waiting time. 

The present simulation for the trapping diffusion model showed that the relaxation function can be fitted by the KWW function only in a limited range, about 2--3 decades of time, although the relaxation time increases with waiting time for both $T$-up and $T$-down protocols.

Meaning of the material time and internal clock are identical to the scaled time introduced in Appendix B to incorporate the relaxation of the FEL. This means that these times must be identical to the time characterizing the delayed response of the FEL.

The present model also provides a clear explanation for the idea of the fictive temperature. The internal temperature introduced in Eqs. (5)--(7) is a parameter that describes the delayed change in the depth of the FEL after a temperature change, although the TNM model introduced $T_{\mathrm{fic}}$ focusing on the delayed response of structure. Therefore, the internal temperature is essentially the same concept as the TNM fictive temperature, that is, the TNM fictive temperature can be understood as a parameter describing the delayed response of the FEL, but focusing only on the structure.

The present result shown in Fig. 8 indicates that the temporal relaxation time can have a minimum as a function of time, which depends on the relaxation time of the FEL and the random distribution of the release rate $W_n$. This behavior will explain the waiting time dependence observed in experiments and computer simulations.

Finally, we comment on future developments of this model. In the present study, we assume a simple exponential function as a delayed response of $T_{\rm{int}}$ (Eq.~(8)). It may be modified by calculating heat transfer in amorphous solids. Differences in heat transfer due to differences in local structures may lead to a position-dependent $T_{\rm_{int}}$, which may result in some physical behavior such as dynamical heterogeneity~\cite{Luo2017-1, Takeda2024}. To combine with experimental data, the present model should be extended to describe the aging phenomena including the memory effect~\cite{Luo2016, Miyamoto2002} and relaxation decoupling~\cite{Luo2017-2, Wojnarowska2021, Soriano2024}.

In conclusion, the time dependence of the temporal relaxation time after the temperature is raised is highly expected to clarify the temperature dependence of the FEL for non-equilibrium systems. 

\begin{acknowledgments}
The authors would like to thank Prof. Koji Fukao for valuable discussion. This work was supported in part by JSPSKAKENHI Grant Number 18K03573. This work used computational resources of Research Center for Computational Science, Okazaki, Japan (Project: 23-IMS-C121, 24-IMS-C116).
\end{acknowledgments}

\appendix
\section{Trapping diffusion and the KWW function}
In numerous non-equilibrium systems, the slow dynamics is said to be well-fitted by the KWW function. Figure 9 shows $\log(-\log \langle F_s (k,t;s)\rangle_s)$ vs $\log t$ plot for the trapping diffusion model. It is observed that the plot becomes a nearly straight line in a limited area about $2\sim3$ decades. Therefore, local behavior may be represented by the KWW function, but it cannot be used to represent the time dependence in the entire time domain.

\section{Trapping random walk}
We consider the trapping diffusion model without trap-rate distribution, i.e. $\epsilon_n=1$ for all $n$, that is, $W_n(t)$'s are the same  which is written as $W(t)$. In this case, we can solve Eq. (9) exactly by introducing a scaled time $\tilde{t}$
\begin{equation}
\tilde{t}(t)=\int_0^t W(t^{\prime})dt^{\prime} .
\end{equation}
Note that this time is identical to the material time~\cite{Odagaki2023}. The solution for SSIF is given by
\begin{equation}
F_s(k,t)=e^{-2(1-\cos{ka})\tilde{t}(t)},
\end{equation}
and, therefore, the two-time correlation function $F_s (k,t^{\prime},t_w )$ is given by
\begin{equation}
F_s(k,t^{\prime},t_w)=e^{-2(1-\cos{ka})[\tilde{t}(t^{\prime}+t_w)-\tilde{t}(t_w)]}.
\end{equation}
It is straightforward to show that the relaxation time of $F_s (k,t^{\prime},t_w )$ as a function of $t^{\prime}$ is an increasing function of $t_w$ for $T$-down protocol and a decreasing function of $t_w$ for $T$-up protocol. 
The temporal relaxation time of $F_s (k,t)$ is readily derived.
\begin{equation}
\tau_{\mathrm{tmp}}(t)w_0=\frac{1}{2(1-\cos{ka})}\exp \bigg\{ \frac{T_g - T_K}{T(t) - T_K} \bigg\}.
\end{equation}
As shown in Fig. 10, this expression explains the numerical result shown in Fig. 5.

\begin{figure}
\includegraphics[width=10cm]{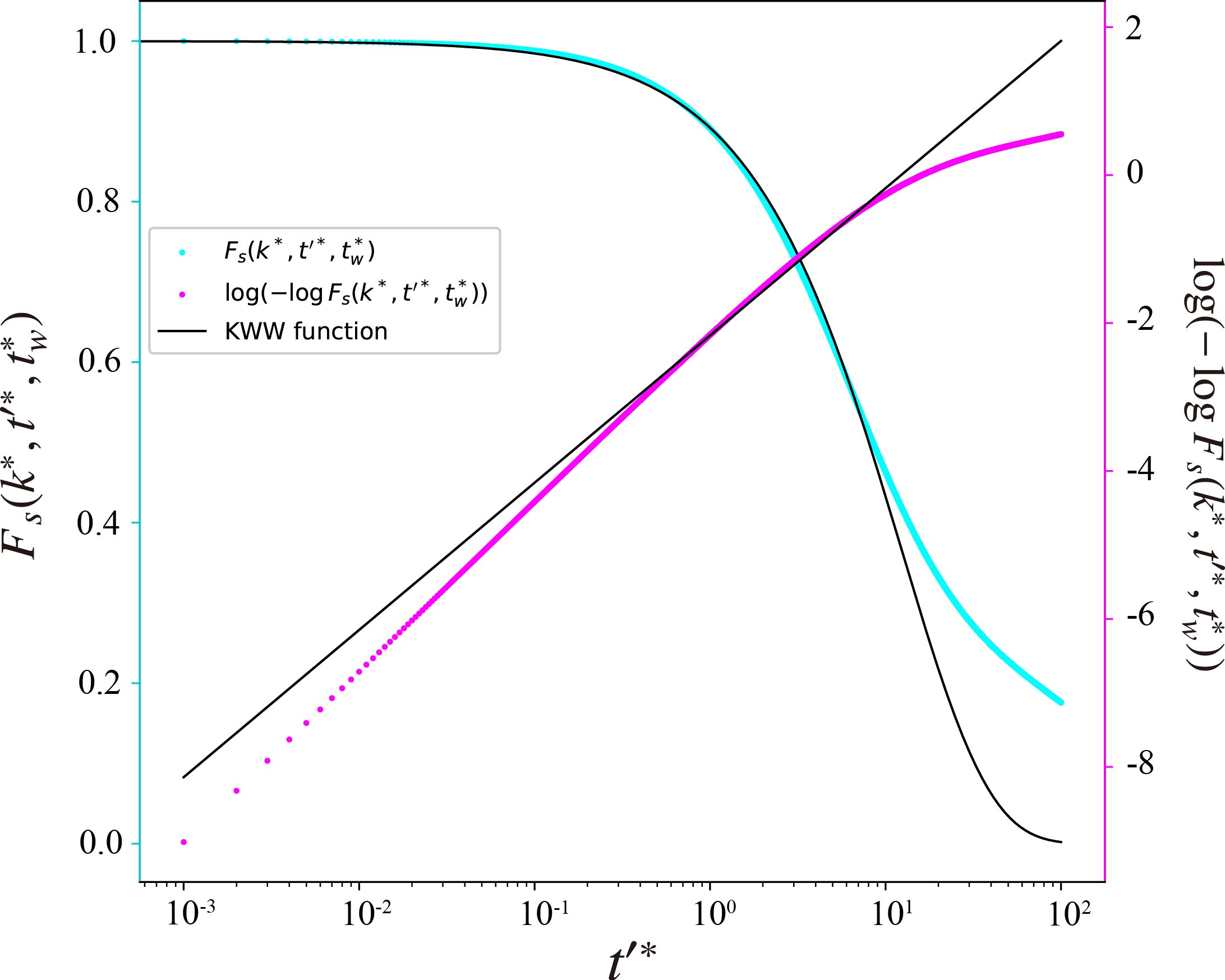}
\caption{SISF for the trapping diffusion model can be fitted by the KWW function in a limited area about two decades for time domain, though it shows slow dynamics.}
\end{figure}

\begin{figure}
\includegraphics[width=10cm]{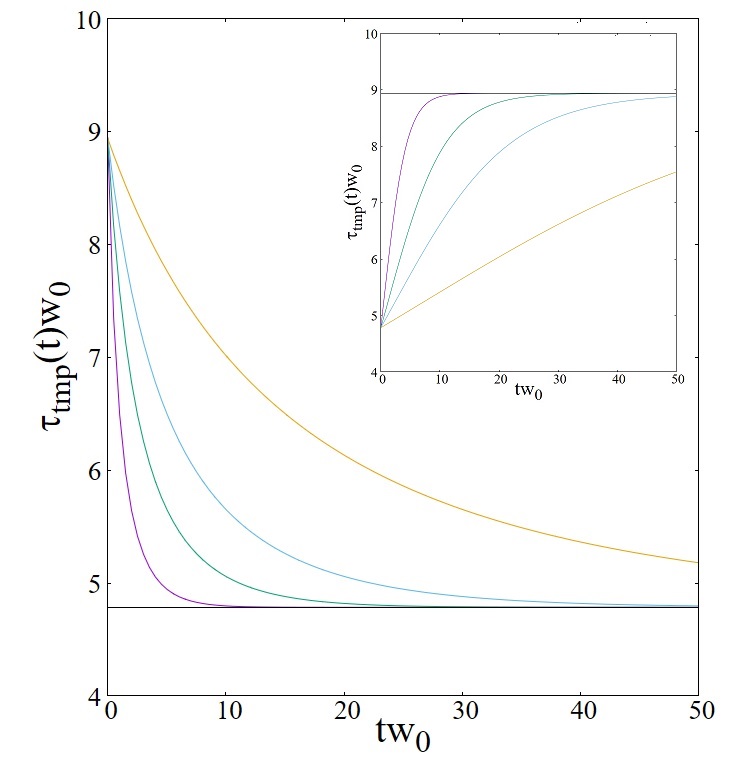}
\caption{The temporal relaxation time of $F_s (k,t)$ for the trapping random walk with the delayed response of the FEL for $T$-up protocol. $\tau_F^*=0,2,5,10$ and 30 from the bottom. Figure 5 agrees with this plot. The inset is for $T$-down protocol for $\tau_F^*=0,2,5,10$ and 30 from the top.}
\end{figure}

\providecommand{\noopsort}[1]{}\providecommand{\singleletter}[1]{#1}%
%

\end{document}